\documentclass[runningheads]{llncs}
\usepackage{graphicx}
\usepackage{amsmath,amssymb} % define this before the line numbering.
\usepackage{color}
\usepackage{multirow}
\usepackage{subcaption}
\usepackage{caption}
\usepackage{array}
\usepackage{booktabs}
\begin{document}
\pagestyle{headings}
\mainmatter

%===========================================================
\title{Blind Image Super-Resolution with Spatial Context Hallucination} % Replace with your title
\author{Dong Huo, Yee-Hong Yang}
%
%\authorrunning{D. Huo and YH. Yang}
\institute{Department of Computing Science\\University of Alberta, Edmonton, Canada\\
\email{\{dhuo,herberty\}@ualberta.ca}}

\maketitle

\begin{abstract}
	Deep convolution neural networks (CNNs) play a critical role in single image super-resolution (SISR) since the amazing improvement of high performance computing. However, most of the super-resolution (SR) methods only focus on recovering bicubic degradation. Reconstructing high-resolution (HR) images from randomly blurred and noisy low-resolution (LR) images is still a challenging problem. In this paper, we propose a novel Spatial Context Hallucination Network (SCHN) for blind super-resolution without knowing the degradation kernel. We find that when the blur kernel is unknown, separate deblurring and super-resolution could limit the performance because of the accumulation of error. Thus, we integrate denoising, deblurring and super-resolution within one framework to avoid such a problem. We train our model on two high quality datasets, DIV2K and Flickr2K. Our method performs better than state-of-the-art methods when input images are corrupted with random blur and noise.
\end{abstract}

\section{Introduction}

Single image super-resolution (SISR) is a fundamental topic in computer vision and has attracted the attention of many researchers for decades. The aim of SISR is to reconstruct a high-resolution (HR) image from its low-resolution (LR) counterpart using only one image. Recently, most of the research in SISR focuses on applying deep learning methods to train an end-to-end model~\cite{dong2015image,haris2018deep,hui2018fast,ledig2017photo,Qiu_2019_ICCV,wang2018esrgan,zhang2018residual} because deep learning methods improve performance significantly. Most of these super-resolution (SR) methods recover HR images from LR images obtained by bicubic downsampling. In particular, HR images are collected from several datasets as training targets and downsampled with bicubic kernel to generate LR training inputs. The procedure is given by 
\begin{gather}
y = x\downarrow_{dowmsample}
\label{eqn:nonblind}
\end{gather}  
where $x\downarrow_{dowmsample}$ represents downsampling HR image $x$ with a downsample kernel (e.g. bicubic kernel), and $y$ is the LR output.

Although these models perform well in recovering bicubic degradation due to downsampling, they are not applicable to more complicated scenarios, e.g.
\begin{gather}
y = (x\circledast k)\downarrow_{dowmsample} + n.
\label{eqn:blind}
\end{gather}
In this case, an image is blurred with a blur kernel $k$ (e.g. a Gaussian blur kernel) before downsampling, and $ \circledast$ represents the convolution operation. $n$ is additive Gaussian noise with noise level $\sigma$. SRMD~\cite{zhang2018learning} stretches the blur kernel and Gaussian noise level to the same size of LR images, and concatenates these three to form the final input. DPSR~\cite{zhang2019deep} regards the deblurring module as a plug-and-play block, and then inputs the deblurred results into the SRResNet~\cite{ledig2017photo} models trained on bicubic degradation. But these models, which reconstruct HR images with known blur kernel and noise level, do not satisfy the condition of blind super-resolution where the explicit degradation model is unknown~\cite{han2010blind}.

Most conventional blind SR algorithms~\cite{begin2004blind,wang2005patch} use optimization instead of deep learning. Michaeli and Irani ~\cite{michaeli2013nonparametric} exploit recurrence of patches across scales and search similar patterns to estimate the blur kernel. However, when large noise is added after downsampling, the estimation accuracy of these models is significantly reduced. Recently, deep learning has been applied to blind SR. For example, SFTMD-IKC~\cite{gu2019blind} extends the work of SRMD and builds an extra subnetwork to learn the blur kernel and exploits spatial feature transform (SFT)~\cite{wang2018recovering} instead of direct concatenation, and upgrades the parameters of IKC iteratively to improve kernel estimation. It learns the blur kernel explicitly which limits the categories of blur kernels and hence, the results can also easily be affected by noise. Besides, SFTMD-IKC contains over 9M parameters which is extremely computational expensive, while the proposed method uses only around 2M (after bypassing parameters that are not used in testing) and is the fastest among all compared SOTA methods. ZSSR~\cite{shocher2018zero} is trained on a single image to take full use of unique internal information. The model is trained in different scales of the input image and exploits the kernel estimation algorithm of Michaeli and Irani ~\cite{michaeli2013nonparametric} when the blur kernel is unknown. The architecture of ZSSR is simple because it is difficult to train a complex model using a single image only. Zhou et al. ~\cite{Zhou_2019_ICCV} estimate the blur kernel from real images with the dark-channel prior~\cite{he2010single} to generate a kernel pool, from which kernels are randomly selected to blur the LR images. Although this method is more general, the performance of the model highly relies on the kernel estimation algorithm~\cite{he2010single} whose outputs are regarded as the ground truth, which may cause the accumulation of error. In addition, it still needs to upsample the LR input to the same size of the HR output which increases computational cost and may introduce extra noise. Most of these methods apply an existing denoising algorithm~\cite{liu2018image,tian2019enhanced,zhang2017beyond} before SR reconstruction to reduce noise. Unfortunately, because these methods are all trained using noisy images, they cannot differentiate between noise-free images and noisy images. Hence, the performance on noise-free LR images is poor because they assume that the input images have noise.

Given the above observations, we propose a new Spatial Context Hallucination Network (SCHN) to combine noise removal and blind super-resolution. Unlike existing SR methods, the proposed method is lightweight and computational efficient in testing by bypassing parameters that are not used during testing (details are shown in Section 4). Our method is an inverse procedure of multiview SR like jittered lens SR~\cite{li2018jittered} and video SR~\cite{tian2018tdan}. In~\cite{li2018jittered} , multiple views of a scene captured by camera jittering differ in only sub-pixel (less than 1 pixel). In~\cite{tian2018tdan}, if the fps of a video is high enough, neighboring frames also differ in sub-pixel. Multiview SR is effective as different views contains more features than a single view. Indeed, our SCH module is to recover the multiview features to enhance the original features. In particular, it learns two offset maps (hallucination maps) that are used to recover the feature maps of neighboring frames in a video, or feature maps of jittered frames from camera jittering, so that we can get more simulated features to perform multiview SR. Each pixel of the feature map is assigned two vectors in the format $(x_{offset}, y_{offset})$, and the value of an offset is less than 1 to make sure that the shifted feature map and the original feature map differ in only sub-pixel. We stacked multiple SCH modules so that we can enhance the feature information by multiview simulation for multiple times. Meanwhile, the shifted feature maps can also compensate the information loss that results from the noise. Since the noise is randomly applied on the image, pixels with large noise level may have neighboring pixels with relative low noise level. In this case, the information of the latter can be shifted to the former which is helpful for image denoising.

The contributions of this work are summarized as follows: (1) We propose a new spatial context hallucination (SCH) module in our network to take full advantage of spatial correlation. To our best knowledge, we are the first to propose such a novel idea. (2) The experiments show that the proposed network, with shallower and thinner architecture, outperforms the state-of-the-art (SOTA) SR methods using the proposed SCH module. (3) Our method integrates denoising, deblurring and upsampling together to solve the blind SR problem and can tolerate more general and random degradation.

\section{Related Work}

The first CNN-based SISR model is the SRCNN\cite{dong2015image}. Its simple architecture shows that it is easy to build an end-to-end CNN model for solving the SR problem. Since then, many CNN-based SR models have been proposed. The VDSR~\cite{kim2016accurate} extends the depth of the SRCNN and adds a long-term residual connection to avoid the vanishing/exploding gradients problem. A pixel-shuffle module of ESPCN~\cite{shi2016real} avoids introducing extra noise from upsampling the LR input into the same size of the HR output. The SRResNet~\cite{ledig2017photo} utilizes both short-term residual blocks and long-term residual connections, followed by multiple sub-pixel modules to gradually upsample the LR input. The architecture of EDSR~\cite{lim2017enhanced} is similar to that of the SRResNet but it removes the batch normalization module to reduce memory usage. The IDN~\cite{hui2018fast} trains a weight for residual connection to distill long-term information. The RDN~\cite{zhang2018residual} combines dense connection and long-term residual connection together to stabilize training. The DBPN~\cite{haris2018deep} not only uses dense connection but also applies upsampling and downsampling alternatively on a single network to preserve HR features inside the network. The MsRN~\cite{gao2019multi} extends the work of EDSR and the RDN to infer multi-scale HR images at the same time. EBRN~\cite{Qiu_2019_ICCV} utilizes multiple block residual modules to restore different frequencies in models of different complexity, which reduces the problem of over-enhancing and over-smoothing. Since one model may not perform best in all images, RankSRGAN~\cite{Zhang_2019_ICCV} integrates SRGAN~\cite{ledig2017photo} and ESRGAN~\cite{wang2018esrgan} by a ranker network to distill the best result from two outputs, and guides the training of SRGAN with a ranking score.

\begin{figure*}[t]
	\begin{center}
		\includegraphics[width=0.75\linewidth]{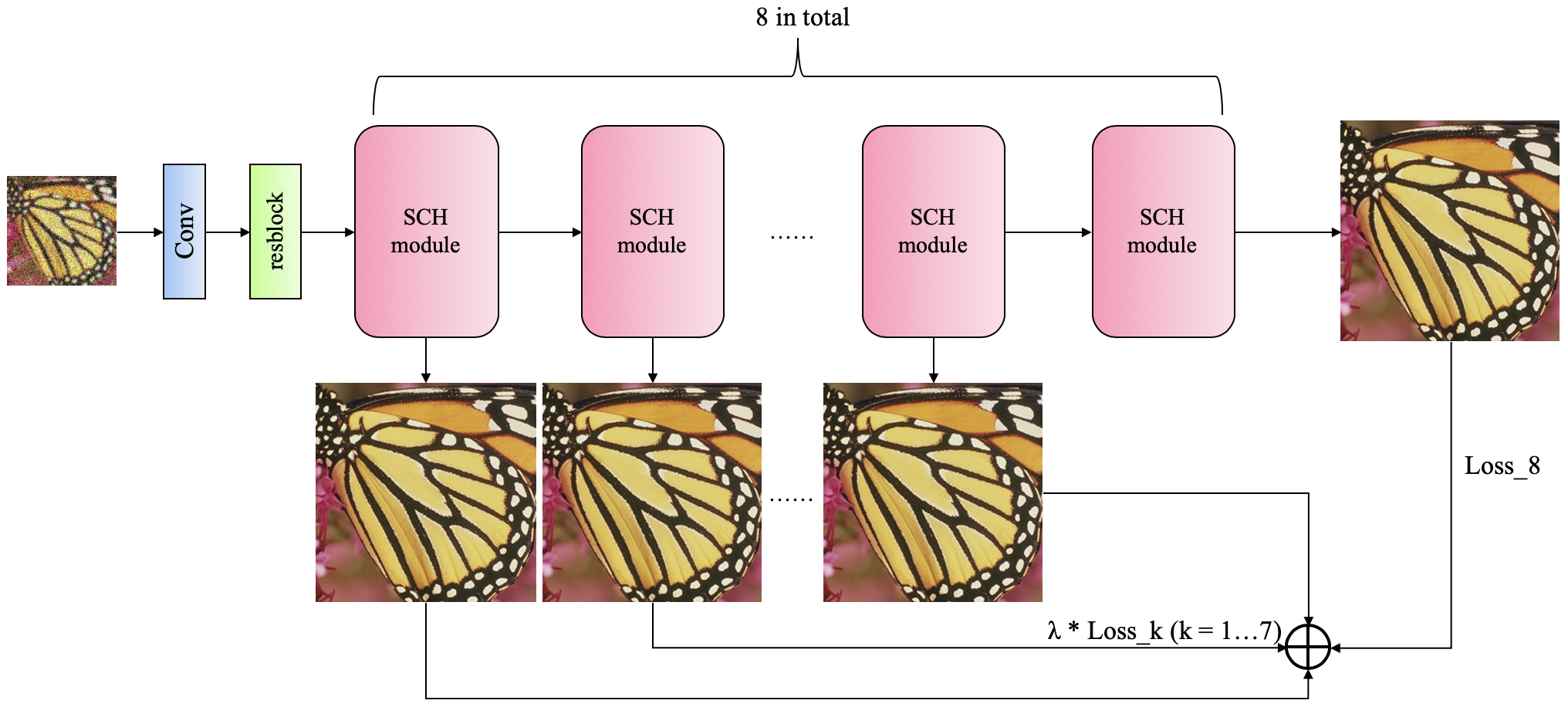}
	\end{center}
	\caption{The network architecture of spatial context hallucination network.}
	\label{fig:architecture}
\end{figure*}
In recent years, the number of published paper in blind SR is still relatively small compared with that of non-blind SR. Some models can be regarded as a hybrid between blind and non-blind SR method. For example, SRMD~\cite{zhang2018learning} assumes that the blur kernel is known and projects it onto a low dimensional vector by Principal Component Analysis (PCA). The kernel vector is concatenated with the $\sigma$ value of noise (noise level) and then, the result is stretched to the same size of the LR input. It uses this extra information as another input to train an SR model. DPSR~\cite{zhang2019deep} utilizes the half quadratic splitting (HQS) algorithm based on the FFT (Fast Fourier Transform) as its deblurring method, which still needs an accurate blur kernel as input. Similar to DPSR, ZSSR~\cite{shocher2018zero} also uses a conventional kernel estimation algorithm~\cite{michaeli2013nonparametric} when it is applied to blind SR, but it is trained on a single image only. In particular, it has eight convolution layers, which is extremely simple compared with most of the SOTA (state-of-the-art) SR networks. The CinCGAN~\cite{yuan2018unsupervised} also exploits the unsupervised strategy in blind SR. Different from ZSSR, it is not trained on LR-HR pairs. However, due to its complicated structure and the ill-posed problem of SR, training CinCGAN is difficult and unstable~\cite{wang2019deep}. SFTMD-IKC~\cite{gu2019blind} splits the network into three parts: a non-blind SR network, a predictor and a corrector of kernel estimation. It converts the stretched blur kernel into two parameters of an SFT module and trains a separate subnetwork to estimate the blur kernel. Then the real kernel input of SFTMD is replaced with the estimated one. Zhou et al.~\cite{Zhou_2019_ICCV} utilize the dark-channel prior~\cite{he2010single} on real images to collect realistic blur kernels, and train a GAN to generate more blur kernels with the same distribution. Then the collected blur kernels are used to generate training data.

SISR is not the only research topic in SR. Li et al.~\cite{li2018jittered} reconstruct HR images from several LR counterparts. In particular, they gather images by jittering the lens of a camera when the shutter is released and images obtained in this way differ in sub-pixel positions only. They then pick pixels one by one from these LR images and insert them into the HR output. Some of the SR methods utilize stereo image pairs as their input. Jeon et al.~\cite{jeon2018enhancing} build a 32-layer network for the stereo SR problems. They first upsample the left and right images using bicubic interpolation and then shift the right image 64 times to build a cost volume as input. The PASSRnet~\cite{wang2019learning} uses matrix multiplication to fuse disparity estimation and SR together. It exploits a residual ASPP block which is designed to enlarge the receptive field and learn multi-scale features in the same layer. 

\begin{figure*}[t]
	\begin{center}
		\includegraphics[width=0.75\linewidth]{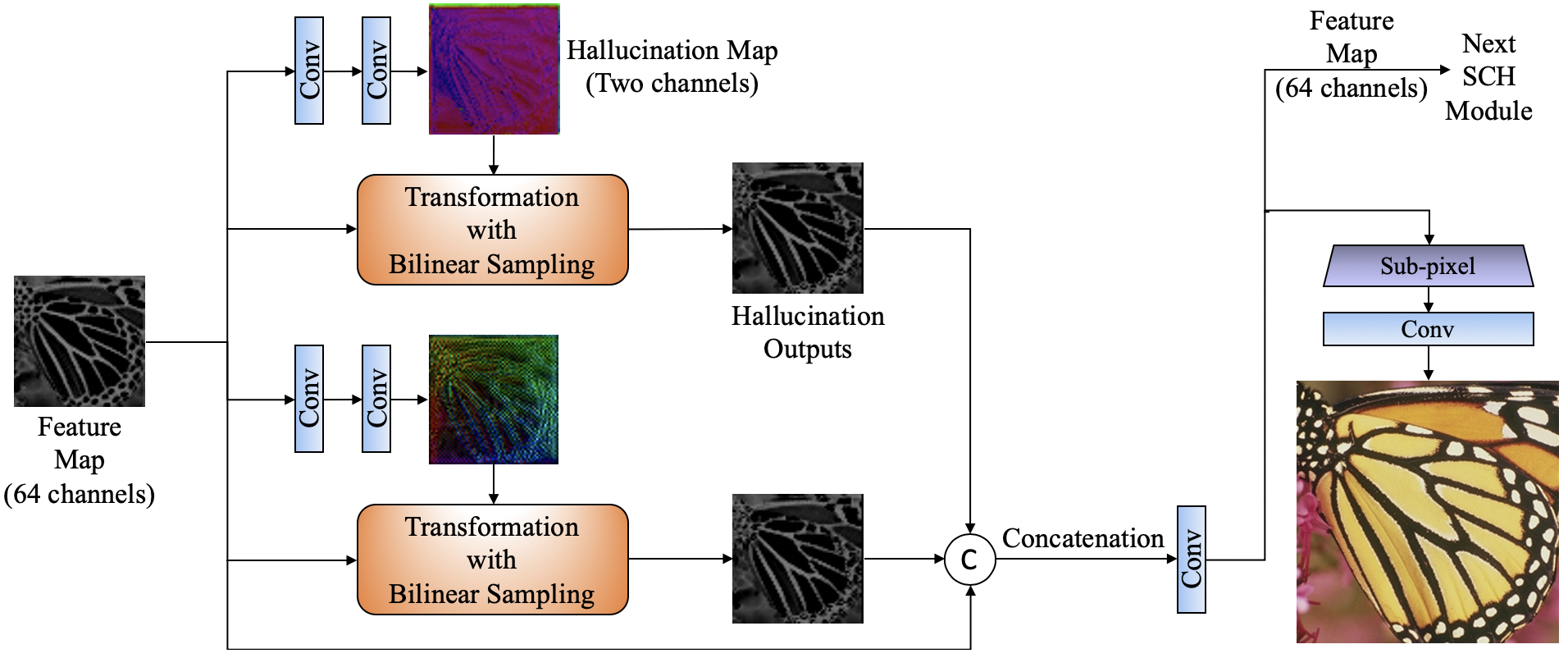}
		\caption{The architecture of the SCH module. The hallucination map is applied to all channels of feature maps. We only display one channel of the feature map before and after ``Transformation with Bilinear Sampling'' for simplification. To visualize the two hallucination maps, we enhance the magnitude of the offsets of each hallucination map. The two channels of each hallucination map are combined and displayed using the convention as that used to display optical flow. Both the input and output feature maps contain 64 channels.}
		\label{fig:shift}
	\end{center}
\end{figure*}

There are also some methods that focus on reconstructing high quality frames from a low-quality video. TDAN~\cite{tian2018tdan} utilizes deformable convolutions~\cite{dai2017deformable} to align the consecutive neighboring LR frames with the reference frame and concatenate the deformed output with feature maps of the reference frame to generate an HR reference frame. EDVR~\cite{wang2019edvr} extends the work of TDAN which uses pyramid and cascading deformable convolutions to exploit the alignment in different scales, and then uses the attention mechanism to fuse the features from different times with different weights. Godard et al.~\cite{godard2018deep} utilize the recurrent architecture to restore all the frames of an image sequence, and address the denoising and SR problems with the same network. PFNL~\cite{Yi_2019_ICCV} exploits a non-local residual block to extract temporal dependencies, and also uses multiple progress fusion resblocks to take full use of spatio-temporal correlations.

In the proposed method, we apply deformable convolutions~\cite{dai2017deformable} to generate new feature maps with sub-pixel offsets in our spatial context hallucination (SCH) module, and use pixel-shuffle convolution~\cite{shi2016real} within each SCH module to enlarge the spatial dimension. Our proposed model achieves the SOTA performance in blind SISR problem. 
\begin{figure*}[t]
	\begin{center}
		\includegraphics[width=0.7\linewidth]{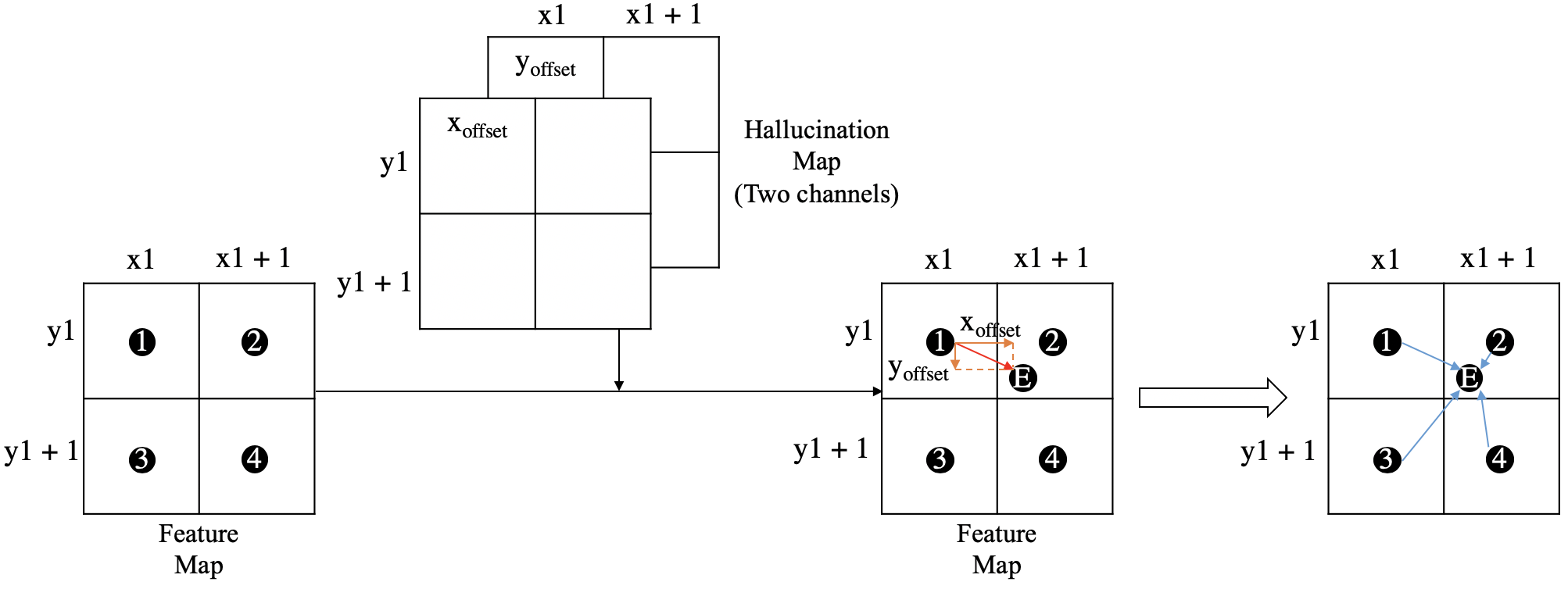}
		\caption{Transformation with bilinear sampling}
		\label{fig:bilinear}
	\end{center}
\end{figure*}
\section{Method}
Our model addresses the problem of blind SR which is formulated as shown in Eqn~\ref{eqn:blind}. SFTMD-IKC~\cite{gu2019blind} focuses on recovering the HR image from an LR input blurred by an isotropic Gaussian blur kernel. In order to make our model more robust, our method recovers LR images blurred by random anisotropic Gaussian blur kernel, which combines Gaussian blur with motion. We also consider removing the additive random Gaussian noise caused by degradation. However, most of the denoising methods~\cite{liu2018image,tian2019enhanced,zhang2017beyond,rudin1992nonlinear} are applicable to noisy images only. As a result, when they are applied to noise-free images, these methods may remove some tiny details or smooth the input inappropriately. To address this problem, our training input includes both noisy and noise-free images mixed together randomly. 

\subsection{Overall Framework}
The proposed spatial context hallucination network (SCHN) is shown in Figure~\ref{fig:architecture}. Our network contains a $3\times3$ convolution layer and a resblock~\cite{he2016deep} followed by 8 stacked SCH modules, and each SCH module generates one high-resolution output. To build the training set for our network, we first utilize random anisotropic Gaussian blur kernels and bicubic downsampling to degrade the HR image set $Z$ and output the noise-free LR image set $Y$. Then we add Gaussian noise with random noise level on $Y$ to generate noisy LR set $X$. Details are given in section 4. To be specific, our random noise level can be 0, and we also randomly skip the blurring process during the dataset generation. 

\begin{figure*}[t]
	\captionsetup[subfigure]{labelformat=empty}
	\begin{center}
		\centering
		\begin{subfigure}{0.385\textwidth}
			\centering
			\includegraphics[width=1\linewidth]{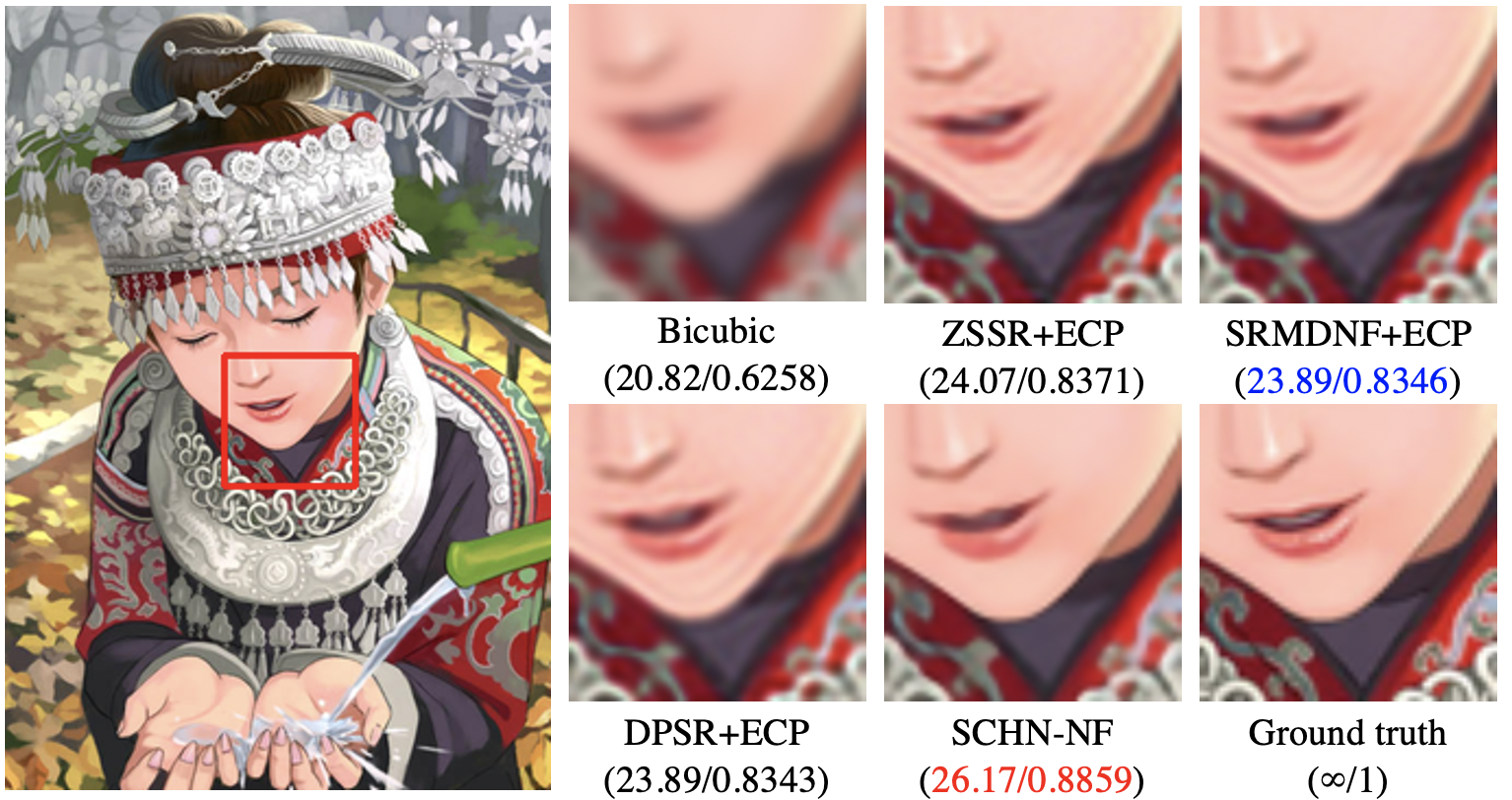}
			\caption{(a) $\sigma$=(1.5, 1.5), sf=2, n=0}
			\label{fig:sub_1}
		\end{subfigure}
		\begin{subfigure}{0.52\textwidth}
			\centering
			\includegraphics[width=1\linewidth]{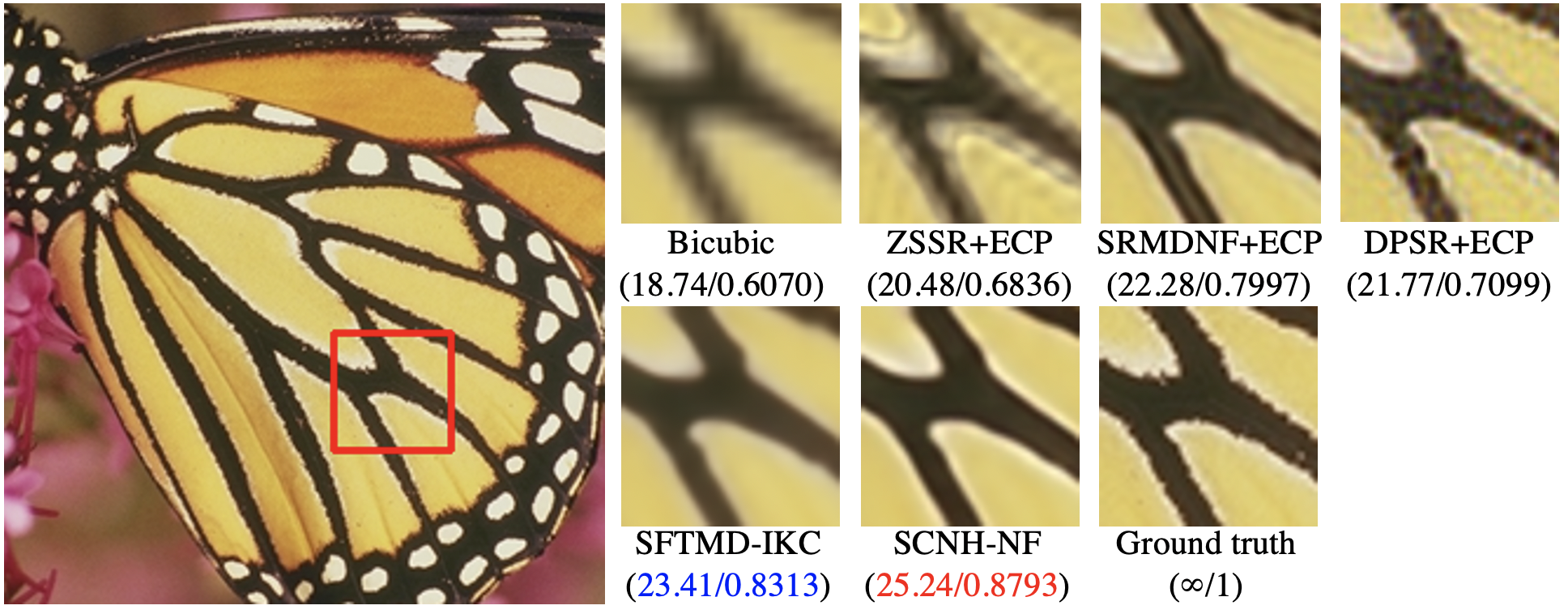}
			\caption{(b) $\sigma$=(2.0, 2.0), sf=4, n=0}
			\label{fig:sub_2}
		\end{subfigure}
		
		\begin{subfigure}{0.455\textwidth}
			\centering
			\includegraphics[width=1.0\linewidth]{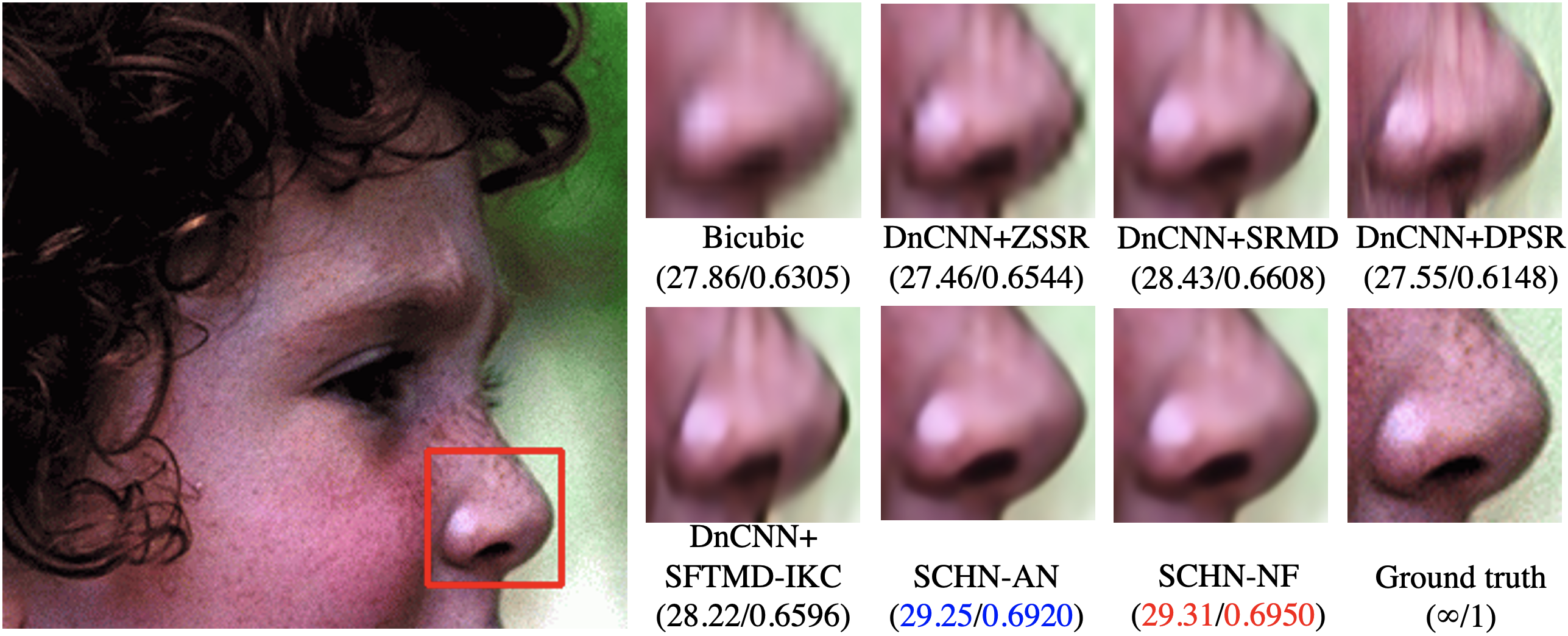}
			\caption{(c) $\sigma$=(0.5, 3.0), sf=4, n=0} 
			\label{fig:sub_3}
		\end{subfigure}
		\begin{subfigure}{0.45\textwidth}
			\centering
			\includegraphics[width=1.0\linewidth]{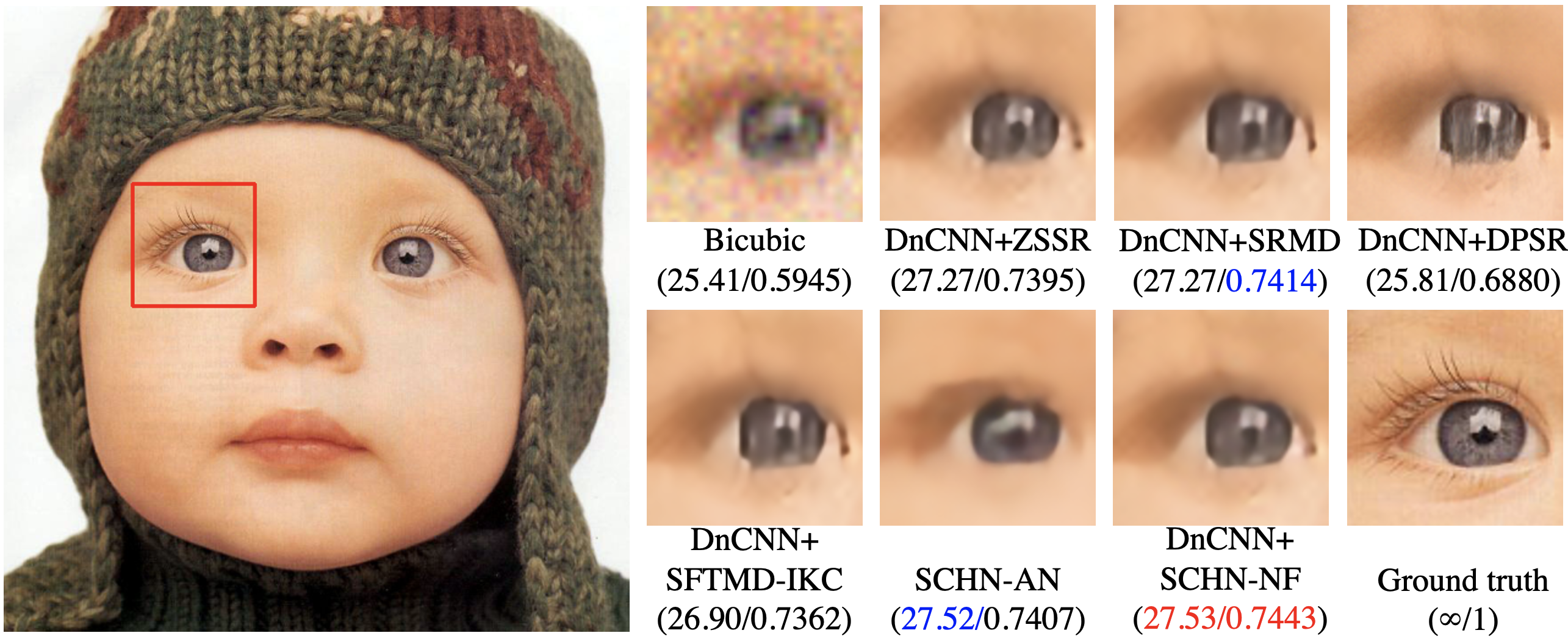}
			\caption{(d) $\sigma$=(0.5, 3.0), sf=4, n=15}  
			\label{fig:sub_4}
		\end{subfigure}
		
		\begin{subfigure}{0.45\textwidth}
			\centering
			\includegraphics[width=1.0\linewidth]{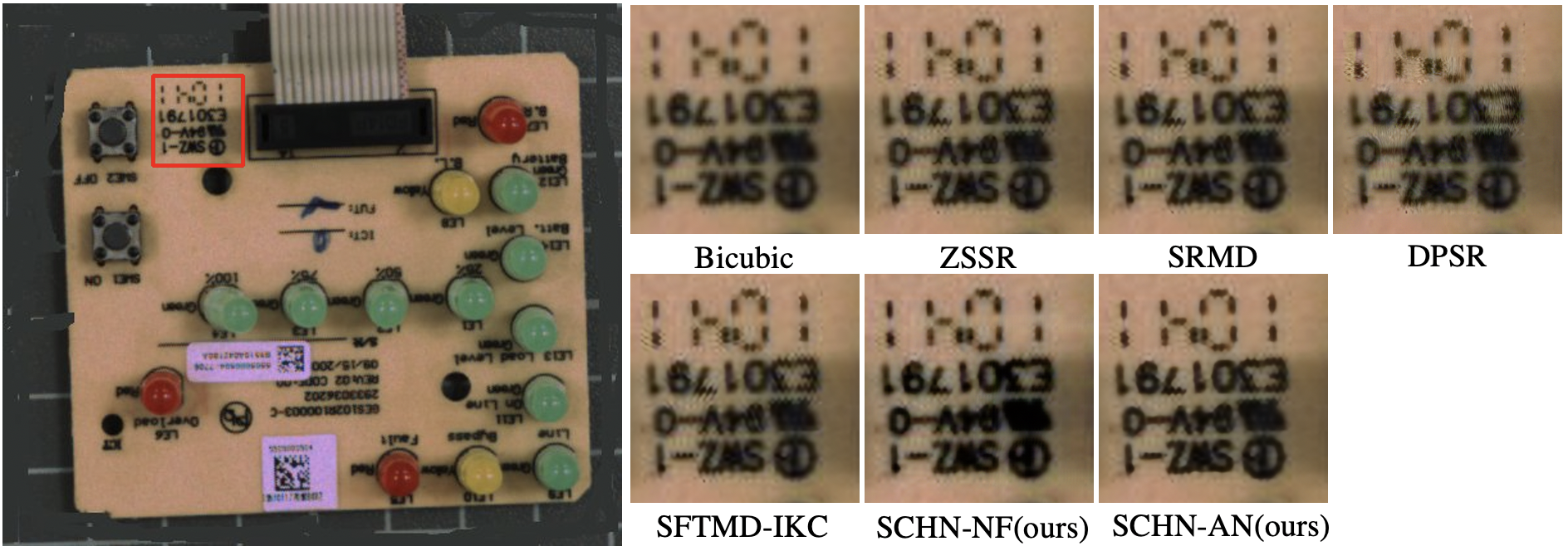}
			\caption{(e) real low-resolution image}  
			\label{fig:sub_5}
		\end{subfigure}
		\begin{subfigure}{0.455\textwidth}
			\centering
			\includegraphics[width=1.0\linewidth]{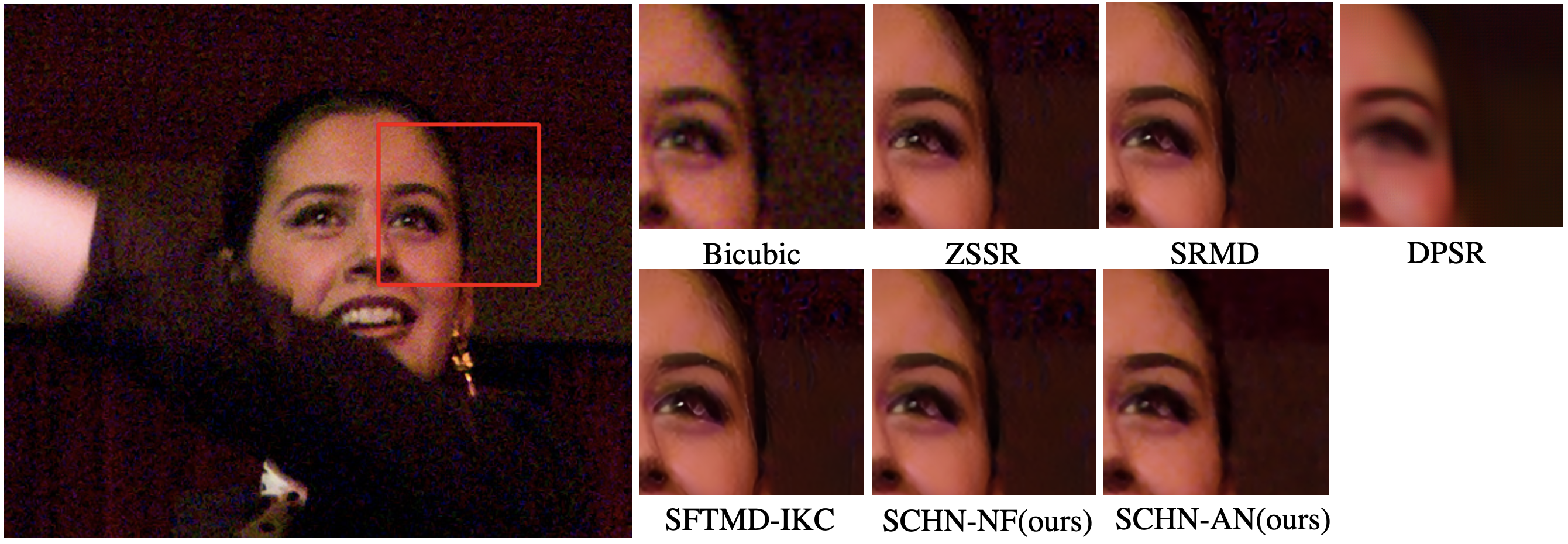}
			\caption{(f) real low-resolution image}
			\label{fig:sub_6}
		\end{subfigure}
		
		\caption{Ouput samples of images on different degradations, and PSNR and SSIM are shown below each of the result. ``$\sigma$" represents the Gaussian blur kernel width; ``sf" represents the scale factor; ``n" represents the noise level. The best results are in \textcolor{red}{red} and the second best in \textcolor{blue}{blue}.}
		\label{fig:iso_visual_comparison_1}
	\end{center}
\end{figure*}
\begin{figure*}[t]
	\captionsetup[subfigure]{labelformat=empty}
	\begin{center}
		\centering
		\includegraphics[width=0.98\linewidth]{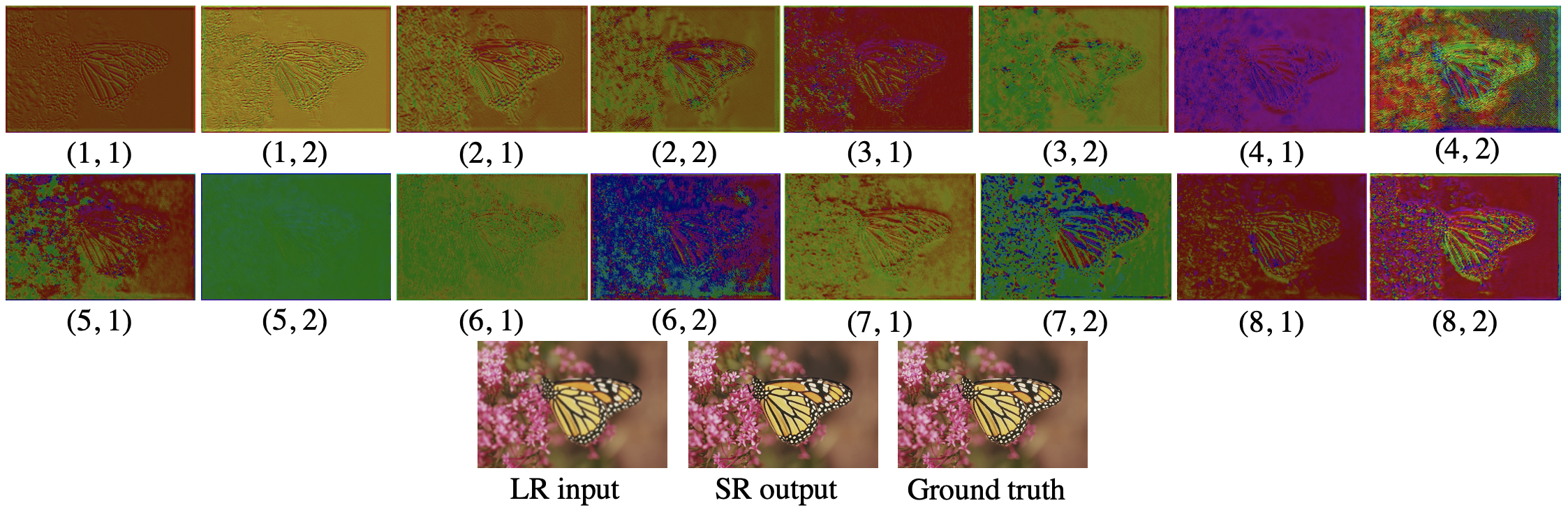}
		\caption{Hallucination maps of SCHN-NF with kernel width 2.0 and scale factor 4. (x, y) represent the y-th hallucination map of the x-th SCH module. Note that we enhance the magnitude of each hallucination map for visualization.}
		\label{fig:hallucination_map}
	\end{center}
\end{figure*}

Similar to the Stacked Hourglass Networks~\cite{newell2016stacked} that output a heatmap and calculate the loss at the end of each hourglass module, we calculate the loss of each high-resolution output. The loss function of each module is given by
\begin{gather}
L^{sr}_j = \frac{1}{N} \sum_{i=1}^{N} \lVert S_j(O_{j - 1}) - z_i\lVert_1, j = 1 ... 8, O_0 = y_i
\label{eqn:single_module_loss}
\end{gather}  
where $z_i \in Z$, $N$ is the batch size, $S_j()$ represents the SCH module and $j$ is the index of a module. $O_{j - 1}$ represents the output of the $(j-1)_{th}$ module. Note that the input of the $1st$ SCH module is the noise-free LR image $y_i \in Y$. Motivated by the work of Zhao et al.~\cite{zhao2015loss}, we use $L1$ loss instead of $L2$ because SSIM and PSNR benefit more from $L1$. The total loss function of our network is given by
\begin{gather}
L^{sr} = \lambda \sum_{j = 1}^{7} L^{sr}_j + L^{sr}_{8},
\label{eqn:total_sr_loss}
\end{gather}  
in which $\lambda$ (we set it as 0.05) is a balancing coefficient to reduce the training conflict of multi-task learning~\cite{sener2018multi}. The output of $S_{8}()$ is selected as the final high resolution output. 

Based on the procedure of Eqn~\ref{eqn:blind}, both the blur kernel $k$ and the downsample kernel $\downarrow_{dowmsample}$ process the HR images by convolution, while the noise component focuses on single pixels. In this case, we integrate the deblurring and SR together, which is different from the CinCGAN~\cite{yuan2018unsupervised} that deblurs and denoises the input simultaneously.  However, we not only want to test the performance of our network on non-linear inverse problem (deblurring and SR), but also attempt to test the compatibility of our network on recovering linear (denoising) and non-linear degradation. The loss function of the former is given in Eqn~\ref{eqn:single_module_loss} and Eqn~\ref{eqn:total_sr_loss}. To test the latter, Eqn~\ref{eqn:single_module_loss} is replaced by
\begin{gather}
L^{sr}_j = \frac{1}{N} \sum_{i=1}^{N} \lVert S_j(O_{j - 1}) - z_i\lVert_1, j = 1 ... 8, O_0 = x_i
\label{eqn:single_module_loss_AN}
\end{gather} 
where the only difference is the noisy input $x_i \in X$. 

\begin{figure*}[t]
	\captionsetup[subfigure]{labelformat=empty}
	\begin{center}
		\begin{subfigure}{0.5\textwidth}
			\centering
			\includegraphics[width=1.0\linewidth]{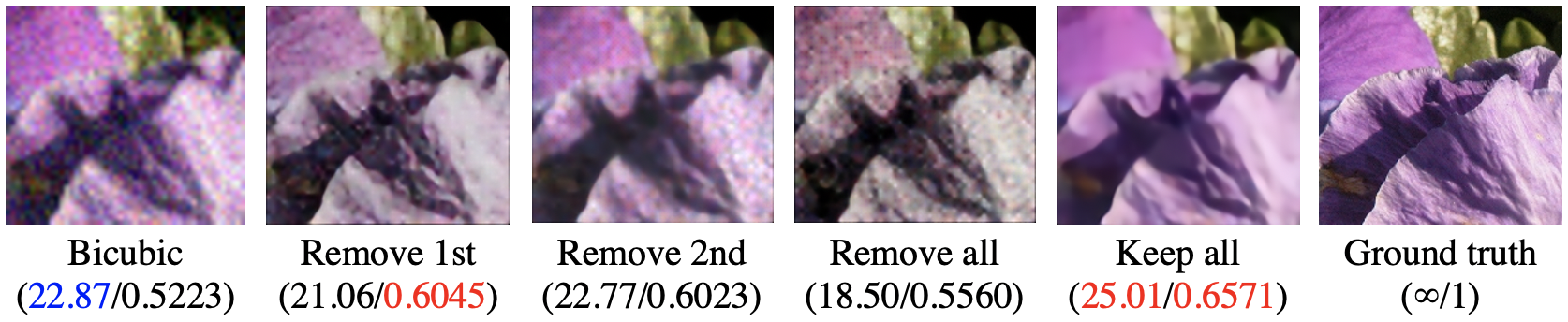}
			\caption{(a) $\sigma$=(1.5, 2.0), sf=4, n=15}
			\label{fig:remove_1}
		\end{subfigure}
		\begin{subfigure}{0.47\textwidth}
			\centering
			\includegraphics[width=1.0\linewidth]{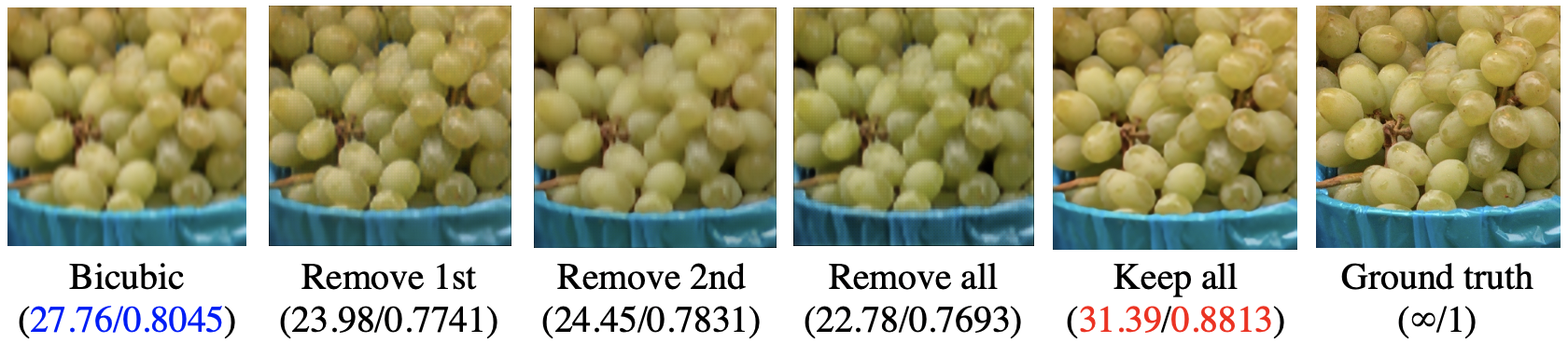}
			\caption{(b) $\sigma$=(0.5, 3.0), sf=4, n=0}
			\label{fig:remove_2}
		\end{subfigure}
	\end{center}
	\caption{``Keep all" represents our original SCHN-AN containing 8 SCH modules with 2 hallucination outputs inside. ``Remove the 1st/2nd" represents that we set the 1st/2nd hallucination output (in all SCH modules) as 0. ``Remove all" represents that we set the both hallucination outputs as 0.}
	\label{fig:remove}
\end{figure*}
\subsection{Spatial Context Hallucination (SCH) Module}
The spatial context hallucination (SCH) module is the core of our network. As we discussed in Section 1, the SCH module is used to simulate camera lens jittering and create multiple pseudo frames, which can be regarded as the inverse procedure of TDAN~\cite{tian2018tdan} and the jittered lens SR~\cite{li2018jittered}. 

We extend the idea of the DCN~\cite{dai2017deformable} which generates only one single offset for each pixel. In our SCH module (shown in Figure~\ref{fig:shift}), the feature map from the last layer are passed through two different branches, and each branch consists of two $3\times3$ convolution layers to generate two offset fields (we call it the hallucination map), one for the x-direction and the other y. The hallucination map, which contains 2 channels, is similar to the optical flow and is applied to the original feature map to generate the offset output by transformation with bilinear sampling. We concatenate these two outputs with the original feature map and the concatenated result is passed to the third $3\times3$ convolution layer followed by a sub-pixel convolution layer~\cite{shi2016real} to enlarge the spatial dimension. There is a Leaky-ReLU activation function after the first and third convolution layers which are not shown in Figure~\ref{fig:shift} explicitly. The output of the second activation function is passed to the next SCH module except the last one. The last convolution layer is used to integrate channels into 3 (RGB). 

The details of transformation are shown in Figure~\ref{fig:bilinear}. The hallucination map contains 2 channels which represent $x_{offset}$ and $y_{offset}$ of each pixel respectively. These offsets are used to obtain the location of the pixel from which the pixel intensity is computed. As shown in Figure~\ref{fig:bilinear}, assume that pixel E is the location of pixel 1 after adding its corresponding offsets in the hallucination map. The intensity at pixel E is computed using bilinear interpolation with pixel 1-4. The resulting intensity is then used to replace the value at pixel 1. More precisely, based on the formula given by Press et al.~\cite{pressnumerical}, 
\begin{equation}
\begin{split}
P(E) = &(x_T + 1 - x) (y_T + 1 - y) P(x_T, y_T)+(x - x_T) (y_T + 1 - y) P(x_T + 1, y_T)\\
+ &(x_T + 1 - x) (y - y_T)P (x_T, y_T + 1)+ (x - x_T) (y - y_T) P(x_T + 1, y_T + 1).
\end{split}
\label{eqn:bilinear}
\end{equation}
where $x_1$ and $y_1$ are the coordinates of pixel 1, $x_{T} = \lfloor x_1 + x_{offset} \rfloor$, $y_{T} = \lfloor y_1 + y_{offset} \rfloor$, $x = x_1 + x_{offset}$ and $y = y_1 + y_{offset}$. Pixels outside of the border are assigned a value of 0. In the above example, pixel 1 gets the value of $P(E)$.

By exploiting the SCH module, features at each pixel are enhanced by shifted features from its neighboring pixels. Details are given by
\begin{equation}
\begin{split}
F(x)_{enhanced} = w_0 F(x)_{original} + \sum_{i=1}^{N}w_i H_i(x)
\end{split}
\label{eqn:enhance}
\end{equation}
in which $x$ is one of the pixel, $F(x)_{original}$ is the original features of $x$, $H_i$ is the $ith$ hallucination outputs and $H_i(x)$ are features shifted to $x$, $F(x)_{enhanced}$ is the features of $x$ after being enhanced, $w$ is the weight which are learned during training. $N$ is the number of hallucination outputs. In this paper, $N=2$ is found to give the best results. Such a strategy is able to feed more information to each pixel, which plays a critical role in recovering high frequency information, e.g. edges and textures. It also helps to compensate the information loss of pixel $x$ that results from the high noise level. Neighboring features with relative low noise level can shift to $x$, which significantly reduce the noise on the $F(x)_{enhanced}$.

We calculate the loss for all SCH modules to limit the range of values in the hallucination maps, which means that no activation function is used after the second convolution layer. In this case, the initial values of hallucination maps could be quite large, and using only one loss function at the end of the network is difficult to constrain all 8 modules. Hence, we use a loss function after each SCH module to limit the output and to stabilize training.

\section{Experiments}
In this section, we first introduce the implementation details of data preprocessing and network training, and then we compare our model with other existing SR methods. We further discuss the ablation study to evaluate the different versions of SCHN. Due to space limitations, more visual comparison on the same datasets are shown in Section 1-3 of the supplementary materials. 

\subsection{Implementation Details}

The training images are gathered from two high quality datasets DIV2K~\cite{agustsson2017ntire} and Flickr2K~\cite{timofte2017ntire}, which consist of 2391 HR images in total. Following Eqn~\ref{eqn:blind}, we synthesize the LR images by blurring and downsampling the HR images, and adding additive Gaussian noise. We first crop the HR images into 256 $\times$ 256 patches with a stride of 240. The LR-HR image pairs are generated during the training procedure. We define a probability for blurring, so that for each input patch, there is a chance to blur it or to keep it unblur. We apply to each image random 15 $\times$ 15 anisotropic Gaussian blur kernel with kernel width in ranges [0.2, 3.0] and [0.2, 4.0] for SR factors 2 and 4, respectively, to the HR patches, and downsample it with bicubic interpolation based on the SR factor. With regard to the noise, there is also a probability to add it or not. A noise level $\sigma$ is randomly selected from a range (0, 50] for both SR factors. We also randomly rotate or flip the images before blurring. In this case, our training set is generated dynamically, which augments the data significantly.

We implement the SCHN in TensorFlow~\cite{abadi2016tensorflow} on a PC with an Nvidia RTX 2080Ti GPU. Our models are optimized using ADAM~\cite{kingma2014adam} with $\beta_1 = 0.9$, $\beta_2 = 0.999$ and a batch size of 4. We initialize the learning rate to $5 \times 10^{-5}$ and decrease it by half every 10 epochs. The training is stopped after 60 epochs. 
\begin{table*}[t]
	\centering
	\tiny
	\caption{The average PSNR and SSIM results of different methods on Set5~\cite{bevilacqua2012low} and Set14~\cite{zeyde2010single} with different scale factors and isotropic kernel width. The size of the blur kernel is fixed at 15 and the downsample method is bicubic interpolation. The results of SFTMD-IKC are obtained from pretrained models on \url{https://github.com/yuanjunchai/IKC}, which provides only models trained in scale factor 4. The best results are in \textcolor{red}{red} and the second best in \textcolor{blue}{blue}.}
	\begin{tabular}{|c|c|c|c|c|c|c|}
		\hline
		\multirow{2}{*}{Method}                                                                                         & \multicolumn{3}{c|}{Scale Factor x2}                                                                                                                                                                                                                                                                     & \multicolumn{3}{c|}{Scale Factor x4}                                                                                                                                                                                                                                                                                       \\ \cline{2-7} 
		& \begin{tabular}[c]{@{}c@{}}Kernel\\ Width\end{tabular} & Set5                                                                                                                   & Set14                                                                                                                  & \begin{tabular}[c]{@{}c@{}}Kernel\\ Width\end{tabular} & Set5                                                                                                                            & Set14                                                                                                                           \\ \hline
		\begin{tabular}[c]{@{}c@{}}Bicubic\\ ZSSR+ECP\\ SRMDNF+ECP\\ DPSR+ECP\\ SFTMD-IKC\\ SCHN-NF (ours)\end{tabular} & 0.5                                                    & \begin{tabular}[c]{@{}c@{}}30.81/0.8969\\ \textcolor{blue}{32.64}/\textcolor{blue}{0.9253}\\ 32.60/0.9250\\ 32.20/0.9226\\ -/-\\ \textcolor{red}{33.70}/\textcolor{red}{0.9310}\end{tabular} & \begin{tabular}[c]{@{}c@{}}27.66/0.8267\\ \textcolor{blue}{28.95}/\textcolor{red}{0.8714}\\ 28.87/\textcolor{blue}{0.8690}\\ 28.53/0.8685\\ -/-\\ \textcolor{red}{29.64}/0.8689\end{tabular} & 0.5                                                    & \begin{tabular}[c]{@{}c@{}}\textcolor{blue}{26.06}/\textcolor{blue}{0.7721}\\ 24.01/0.7285\\ 23.78/0.7413\\ 19.31/0.4856\\ 22.07/0.6745\\ \textcolor{red}{27.81}/\textcolor{red}{0.8339}\end{tabular} & \begin{tabular}[c]{@{}c@{}}\textcolor{blue}{23.60}/\textcolor{blue}{0.6661}\\ 21.94/0.6417\\ 21.15/0.6309\\ 17.07/0.3593\\ 19.63/0.5991\\ \textcolor{red}{24.79}/\textcolor{red}{0.7056}\end{tabular} \\ \hline
		\begin{tabular}[c]{@{}c@{}}Bicubic\\ ZSSR+ECP\\ SRMDNF+ECP\\ DPSR+ECP\\ SFTMD-IKC\\ SCHN-NF (ours)\end{tabular} & 1.0                                                    & \begin{tabular}[c]{@{}c@{}}28.88/0.8543\\ \textcolor{blue}{32.20}/0.9149\\ 32.08/0.9139\\ 32.08/\textcolor{blue}{0.9150}\\ -/-\\ \textcolor{red}{33.64}/\textcolor{red}{0.9305}\end{tabular} & \begin{tabular}[c]{@{}c@{}}26.15/0.7661\\ 28.74/0.8486\\ \textcolor{blue}{28.95}/0.8489\\ 28.76/\textcolor{blue}{0.8510}\\ -/-\\ \textcolor{red}{29.61}/\textcolor{red}{0.8600}\end{tabular} & 1.0                                                    & \begin{tabular}[c]{@{}c@{}}25.99/0.7617\\ 26.36/0.8050\\ \textcolor{blue}{27.28}/\textcolor{blue}{0.8230}\\ 22.75/0.6382\\ 26.38/0.8064\\ \textcolor{red}{28.69}/\textcolor{red}{0.8442}\end{tabular} & \begin{tabular}[c]{@{}c@{}}23.79/0.6555\\ 24.27/0.6962\\ \textcolor{blue}{24.29}/\textcolor{blue}{0.7065}\\ 20.08/0.4976\\ 23.12/0.6929\\ \textcolor{red}{25.58}/\textcolor{red}{0.7129}\end{tabular} \\ \hline
		\begin{tabular}[c]{@{}c@{}}Bicubic\\ ZSSR+ECP\\ SRMDNF+ECP\\ DPSR+ECP\\ SFTMD-IKC\\ SCHN-NF (ours)\end{tabular} & 1.5                                                    & \begin{tabular}[c]{@{}c@{}}27.11/0.8020\\ \textcolor{blue}{28.44}/0.8474\\ 28.35/\textcolor{blue}{0.8557}\\ 28.36/0.8412\\ -/-\\ \textcolor{red}{32.98}/\textcolor{red}{0.9255}\end{tabular} & \begin{tabular}[c]{@{}c@{}}24.76/0.7001\\ 25.81/\textcolor{blue}{0.8032}\\ \textcolor{blue}{25.97}/0.8019\\ 25.76/0.7497\\ -/-\\ \textcolor{red}{29.31}/\textcolor{red}{0.8596}\end{tabular} & 2.0                                                    & \begin{tabular}[c]{@{}c@{}}24.65/0.7057\\ 26.23/0.8032\\ 27.09/0.7998\\ 26.45/0.7687\\ \textcolor{blue}{28.68}/\textcolor{blue}{0.8321}\\ \textcolor{red}{29.08}/\textcolor{red}{0.8480}\end{tabular} & \begin{tabular}[c]{@{}c@{}}22.89/0.6003\\ 24.34/0.6662\\ 24.89/0.6978\\ 24.25/0.6621\\ \textcolor{blue}{25.56}/\textcolor{blue}{0.7082}\\ \textcolor{red}{25.91}/\textcolor{red}{0.7197}\end{tabular} \\ \hline
		\begin{tabular}[c]{@{}c@{}}Bicubic\\ ZSSR+ECP\\ SRMDNF+ECP\\ DPSR+ECP\\ SFTMD-IKC\\ SCHN-NF (ours)\end{tabular} & 2.0                                                    & \begin{tabular}[c]{@{}c@{}}25.72/0.7509\\ \textcolor{blue}{26.36}/\textcolor{blue}{0.8023}\\ 26.33/0.7986\\ 26.33/0.7746\\ -/-\\ \textcolor{red}{32.29}/\textcolor{red}{0.9190}\end{tabular} & \begin{tabular}[c]{@{}c@{}}23.70/0.6441\\ \textcolor{blue}{24.40}/\textcolor{blue}{0.7371}\\ 24.37/0.7350\\ 24.34/0.7356\\ -/-\\ \textcolor{red}{28.70}/\textcolor{red}{0.8451}\end{tabular} & 3.0                                                    & \begin{tabular}[c]{@{}c@{}}23.36/0.6479\\ 24.02/0.7371\\ 24.40/0.7217\\ 24.44/0.7088\\ \textcolor{blue}{28.45}/\textcolor{blue}{0.8194}\\ \textcolor{red}{28.70}/\textcolor{red}{0.8369}\end{tabular} & \begin{tabular}[c]{@{}c@{}}21.94/0.5501\\ 22.93/0.6190\\ 23.07/0.6390\\ 23.04/0.6251\\ \textcolor{red}{26.04}/\textcolor{red}{0.7178}\\ \textcolor{blue}{25.62}/\textcolor{blue}{0.7046}\end{tabular} \\ \hline
	\end{tabular}
	\label{tab:iso}
\end{table*}
\subsection{Experiments on Isotropic Blur Kernel}
In order to evaluate the performance of our model on the non-linear inverse problem, we first use images blurred with isotropic Gaussian blur kernel and downsampled with bicubic interpolation. Note that we calculate the PSNR and SSIM on the RGB channels instead of only the Y channel in YCbCr color space as we not only want to recover the luminance information but also the chrominance.

Table~\ref{tab:iso} shows the PSNR and SSIM results on two standard benchmark datasets: Set5~\cite{bevilacqua2012low} and Set14~\cite{zeyde2010single} with different scale factors and kernel widths. We compare our proposed SCHN with bicubic interpolation and four SOTA methods including ZSSR~\cite{shocher2018zero}, SRMDNF (noise-free version of SRMD)~\cite{zhang2018learning}, DPSR~\cite{zhang2019deep} and SFTMD-IKC~\cite{gu2019blind}. SCHN-NF represents the SCHN network trained on noise-free blurry images. As ZSSR, SRMDNF and DPSR are all non-blind SR methods which need the blur kernel, we apply extreme channels prior (ECP)~\cite{yan2017image} to estimate the blur kernel after the SR procedure. As one can see, SCHN-NF achieves the best performance in most of the cases. With unknown kernel width, the ZSSR achieves slightly better results than the SRMDNF at small kernel but worse when the width grows. Since a large scale factor and a small kernel width can create more artifacts, the performance of DPSR drops significantly with kernel width 0.5 and scale factor 4. Although the SFTMD-IKC achieves good performance for large kernel width, it does not even outperform bicubic interpolation for small ones, which is the drawback of explicitly estimating the kernel width with IKC, and splitting deblurring (IKC) and SR (SFTMD) may cause the accumulation of error which further impacts the results. 

The visual comparisons are shown in Figure~\ref{fig:sub_1} and~\ref{fig:sub_2}. The ringing effect is obvious for ZSSR, SRMDNF and DPSR with kernel width 2, and even worse for ZSSR with kernel width 4. Besides, the DPSR creates more artifacts than others because of over-enhancing, and the edges of SRMDNF are over-smoothed. Compared with SFTMD-IKC which reduces the intensity of the image because of over-smoothing, our SCHN-NF not only generates sharper edges without artifacts but also preserves the intensity of the images.

We also display the hallucination maps of our SCHN-NF in Figure~\ref{fig:hallucination_map}. The two channels of each hallucination map are combined and displayed using the convention as that used to display optical flow. (x, y) represent the y-th hallucination map of the x-th
SCH module. We can see that the offsets on the edges are obviously different from that of the flatten regions, which means the hallucination maps can help to restore the high frequency features by feeding more information to the edges from multiple directions.

\begin{table*}[t]
	\centering
	\tiny
	\caption{The average PSNR and SSIM results of different methods on Set5~\cite{bevilacqua2012low} with scale factor 4 and different anisotropic kernel width. The size of blur kernel is fixed to 15 and the downsample method is fixed to bicubic interpolation. Note that before the SR procedure, DnCNN are used on all noise levels including 0. The best results are in \textcolor{red}{red} and the second best in \textcolor{blue}{blue}.}
	\begin{tabular}{|c|c|c|c|c|}
		\hline
		\multirow{2}{*}{Method}                                                                                                                                    & \multirow{2}{*}{\begin{tabular}[c]{@{}c@{}}Noise\\ Level\end{tabular}} & \multicolumn{3}{c|}{Kernel Width ($\sigma_x$, $\sigma_y$)}                                                                                                                                                                                                                                                                                                                                                                                                                                                   \\ \cline{3-5} 
		&                                                                        & (0.5, 3.0)                                                                                                                                                    & (1.0, 2.5)                                                                                                                                                    & (1.5, 2.0)                                                                                                                                                    \\ \hline
		\begin{tabular}[c]{@{}c@{}}Bicubic\\ DnCNN+ZSSR\\ DnCNN+SRMD\\ DnCNN+DPSR\\ DnCNN+SFTMD-IKC\\ SCHN-NF (ours)\\ SCHN-AN (ours)\\ DnCNN+SCHN-NF\end{tabular} & 0                                                                      & \begin{tabular}[c]{@{}c@{}}24.67/0.7093\\ 25.18/0.7435\\ 25.83/0.7683\\ 24.52/0.7090\\ 25.13/0.7568\\ \textcolor{red}{28.28}/\textcolor{red}{0.8385}\\ \textcolor{blue}{28.12}/\textcolor{blue}{0.8317}\\ 27.73/0.8149\end{tabular} & \begin{tabular}[c]{@{}c@{}}24.96/0.7181\\ 26.30/0.7691\\ 27.16/0.8044\\ 25.80/0.7480\\ 26.89/0.8037\\ \textcolor{red}{28.89}/\textcolor{red}{0.8467}\\ \textcolor{blue}{28.66}/\textcolor{blue}{0.8390}\\ 28.45/0.8288\end{tabular} & \begin{tabular}[c]{@{}c@{}}25.05/0.7218\\ 26.80/0.7858\\ 28.22/0.8258\\ 26.90/0.7789\\ 28.69/0.8366\\ \textcolor{red}{29.09}/\textcolor{red}{0.8486}\\ \textcolor{blue}{28.81}/\textcolor{blue}{0.8400}\\ 28.76/0.8342\end{tabular} \\ \hline
		\begin{tabular}[c]{@{}c@{}}Bicubic\\ DnCNN+ZSSR\\ DnCNN+SRMD\\ DnCNN+DPSR\\ DnCNN+SFTMD-IKC\\ SCHN-NF (ours)\\ SCHN-AN (ours)\\ DnCNN+SCHN-NF\end{tabular} & 15                                                                     & \begin{tabular}[c]{@{}c@{}}22.90/0.5541\\ 23.92/0.6745\\ 24.36/0.6919\\ 23.47/0.6464\\ 23.96/0.6836\\ 23.13/0.5639\\ \textcolor{red}{25.17}/\textcolor{red}{0.7220}\\ \textcolor{blue}{24.95}/\textcolor{blue}{0.7120}\end{tabular} & \begin{tabular}[c]{@{}c@{}}23.10/0.5623\\ 24.43/0.6844\\ 24.99/0.7088\\ 24.13/0.6646\\ 24.72/0.7036\\ 23.40/0.5746\\ \textcolor{red}{25.47}/\textcolor{red}{0.7286}\\ \textcolor{blue}{25.44}/\textcolor{blue}{0.7254}\end{tabular} & \begin{tabular}[c]{@{}c@{}}23.17/0.5648\\ 24.61/0.6919\\ 25.30/0.7212\\ 24.54/0.6792\\ 25.23/0.7178\\ 23.49/0.5754\\ \textcolor{blue}{25.63}/\textcolor{blue}{0.7328}\\ \textcolor{red}{25.71}/\textcolor{red}{0.7334}\end{tabular} \\ \hline
		\begin{tabular}[c]{@{}c@{}}Bicubic\\ DnCNN+ZSSR\\ DnCNN+SRMD\\ DnCNN+DPSR\\ DnCNN+SFTMD-IKC\\ SCHN-NF (ours)\\ SCHN-AN (ours)\\ DnCNN+SCHN-NF\end{tabular} & 30                                                                     & \begin{tabular}[c]{@{}c@{}}20.21/0.3931\\ 22.85/0.6227\\ 23.16/0.6428\\ 22.30/0.5859\\ 22.80/0.6322\\ 20.04/0.3859\\ \textcolor{red}{23.64}/\textcolor{red}{0.6666}\\ \textcolor{blue}{23.47}/\textcolor{blue}{0.6556}\end{tabular} & \begin{tabular}[c]{@{}c@{}}20.35/0.3999\\ 23.14/0.6335\\ 23.53/0.6544\\ 22.79/0.6057\\ 23.21/0.6454\\ 20.15/0.3912\\\textcolor{red}{ 23.83}/\textcolor{red}{0.6734}\\ \textcolor{blue}{23.77}/\textcolor{blue}{0.6658}\end{tabular} & \begin{tabular}[c]{@{}c@{}}20.39/0.4029\\ 23.31/0.6392\\ 23.74/0.6623\\ 22.93/0.6084\\ 23.46/0.6545\\ 20.16/0.3926\\ \textcolor{blue}{23.87}/\textcolor{red}{0.6734}\\ \textcolor{red}{23.90}/\textcolor{blue}{0.6715}\end{tabular} \\ \hline
		\begin{tabular}[c]{@{}c@{}}Bicubic\\ DnCNN+ZSSR\\ DnCNN+SRMD\\ DnCNN+DPSR\\ DnCNN+SFTMD-IKC\\ SCHN-NF (ours)\\ SCHN-AN (ours)\\ DnCNN+SCHN-NF\end{tabular} & 45                                                                     & \begin{tabular}[c]{@{}c@{}}17.98/0.2929\\ 22.04/0.5891\\ 22.19/0.6053\\ 21.48/0.5459\\ 21.84/0.5935\\ 17.61/0.2763\\ \textcolor{red}{22.45}/\textcolor{red}{0.6190}\\ \textcolor{blue}{22.43}/\textcolor{blue}{0.6178}\end{tabular} & \begin{tabular}[c]{@{}c@{}}18.10/0.2975\\ 22.17/0.5966\\ 22.40/0.6138\\ 21.75/0.5552\\ 22.06/0.6033\\ 17.67/0.2811\\ \textcolor{red}{22.63}/\textcolor{red}{0.6247}\\ \textcolor{blue}{22.56}/\textcolor{blue}{0.6239}\end{tabular} & \begin{tabular}[c]{@{}c@{}}18.10/0.2953\\ 22.31/0.6041\\ 22.58/0.6200\\ 21.96/0.5626\\ 22.27/0.6083\\ 17.66/0.2832\\ \textcolor{red}{22.73}/\textcolor{red}{0.6315}\\ \textcolor{blue}{22.72}/\textcolor{blue}{0.6301}\end{tabular} \\ \hline
	\end{tabular}
	\label{tab:aniso}
\end{table*}
\subsection{Experiments on General Degradation}
To evaluate the compatibility of our network on recovering linear (denoising) and non-linear degradation, we also compare with ZSSR~\cite{shocher2018zero}, SRMD~\cite{zhang2018learning}, DPSR~\cite{zhang2019deep} and SFTMD-IKC~\cite{gu2019blind} on random anisotropic Gaussian blur kernels with Gaussian noise. SCHN-AN represents the SCHN network trained on blurry images with additive noise. Although SRMD and DPSR are also designed for denoising, they need a predefined noise level. Thus, we apply DnCNN (blind denoising version)~\cite{zhang2017beyond} to these four methods for image denoising before the SR procedure. Different from the experiments on isotropic blur kernel, we do not use extra deblurring methods because the DnCNN deblurs the images along with denoising, and extra deblurring actually degrade the performance. We also compare with SCHN-NF without any denoising methods to show the effect of noise.

Table~\ref{tab:aniso} shows the experimental results on anisotropic Gaussian blur kernel with scale factor 4 and Gaussian noise. We select different kernel width pairs ($\sigma_x$, $\sigma_y$) for each of the methods, and add different levels of noise. As one can see, ZSSR, SRMD, DPSR and SFTMD-IKC are all sensitive to the difference of $\sigma_x$ and $\sigma_y$ when the noise level is 0. For the kernel width pair (0.5, 3.0), these methods perform only slightly better than the bicubic interpolation. With non-zero noise level, these methods become more stable because the denoising procedure also recovers some details as deblurring, but still cannot outperform our methods. The performance of SCHN-NF is better than SCHN-AN and DnCNN+SCHN-NF on zero noise level, because the denoising procedure may remove some tiny features which are regarded as noise by mistake. However, our SCHN-AN still outperforms DnCNN+SCHN-NF in most cases which shows that our SCHN-AN works better in denoising and can protect more features from being removed.

As shown in Figure~\ref{fig:sub_3} and~\ref{fig:sub_4} , the results of our methods far surpass that of others when the difference of $\sigma_x$ and $\sigma_y$ is large  which show the robustness of our model. ZSSR, SRMD, DPSR and SFTMD-IKC over-enhance edges which appear as deformation. Although the DnCNN+SCHN-NF outperforms the SCHN-AN in this case, SCHN-AN can generate smoother edges with less artifacts.

\subsection{Experiments on Real Images}
In addition to the synthetic data, we also test our SCHN on real images. We only display the visual comparison (shown in Figure~\ref{fig:sub_5} and~\ref{fig:sub_6}) since there is no ground truth. The testing image is provided by Lebrun et al.~\cite{lebrun2015noise}. As one can see, the result of bicubic interpolation is blurry and noisy which is expected. All of the other 4 methods generate large artifacts especially on straight lines and cannot remove all of the noise on the flatten regions. Our method enhances the edges on the image and remove more noise than others. The experimental results show that our method is more suitable for the real image SR problem.
\begin{table}[t]
	\centering
	\tiny
	\caption{Comparison of efficiency. Note that we do not collect the testing time of SRMD-NF because we test it on a CPU. We still use the ECP~\cite{yan2017image} for blur kernel estimation but we ignore its running time.}
	\begin{tabular}{|c|c|c|c|}
		\hline
		Methods    &  Parameters& Testing Time & PSNR/SSIM    \\ \hline
		ZSSR       & \textcolor{red}{0.22M}                                                            & 37.53s                                                 & 30.48/0.8464 \\ \hline
		SRMDNF     & \textcolor{blue}{1.60M}                                                            & -                                                      & \textcolor{blue}{30.70}/\textcolor{blue}{0.8531} \\ \hline
		DPSR       & 3.49M                                                            & \textcolor{blue}{2.18s}                                                & 24.44/0.6386 \\ \hline
		SFTMD-IKC  & 9.05M                                                            & 7.88s                                                  & 30.00/0.8461 \\ \hline
		SCHN-NF (ours)       & 6.31M                                                            & \textcolor{red}{1.15s}                                                  & \textcolor{red}{31.36}/\textcolor{red}{0.8604} \\ \hline
	\end{tabular}
	\label{tab:efficiency}
\end{table}
\subsection{Comparison of Efficiency}
We also compare the parameters and testing time of our SCHN with others. Since the architecture of SCHN-NF is the same as that of SCHN-AN, we only display the efficiency of SCHN-NF. The testing time is collected by using ``baby.png” of Set5~\cite{bevilacqua2012low} on an Nvidia RTX 2080ti with isotropic kernel width 1.0 and scale factor 4. As shown in Table~\ref{tab:efficiency}, ZSSR has the smallest number of parameters as it contains only 8 convolution layers, while SFTMD-IKC has the largest number of parameters as it contains three different parts including a non-blind SR network (SFTMD), a predictor and a corrector for kernel estimation (IKC). The testing time of ZSSR is much longer than that of others because it uses unsupervised learning and has to train a new model for each testing image. Although SFTMD-IKC uses supervised learning, it needs multiple iterations to refine the high resolution outputs which is the same as the DPSR. Our method achieves the best performance on this image with the lowest run time. Note that we do not remove those parameters which are not used in the testing procedure. Otherwise, our speed would be faster which takes only 0.69s with 2.16M parameters.
\begin{table}[t]
	\centering
	\tiny
	\caption{The average PSNR and SSIM results of SCHN-NF with different number of hallucination maps and SCH modules on T91~\cite{yang2010image} with scale factor 4 and kernel width 2.0. The best results are in \textcolor{red}{red} and the second best in \textcolor{blue}{blue}.}
	\begin{tabular}{|c|c|c|c|c|c|c|}
		\hline
		\multicolumn{2}{|c|}{\multirow{2}{*}{}}                                           & \multicolumn{5}{c|}{SCH modules}                                       \\ \cline{3-7} 
		\multicolumn{2}{|c|}{}                                                            & 0          & 1            & 4            & 8            & 12           \\ \hline
		\multirow{4}{*}{\begin{tabular}[c]{@{}c@{}}Hallucination\\ maps\end{tabular}} & 0 & 26.21/7262 & 26.49/0.7388 & 26.77/0.7480 & 26.82/0.7496 & 26.98/0.7586 \\ \cline{2-7} 
		& 1 & -/-        & 26.63/0.7457 & 27.51/0.7752 & 27.58/0.7759 & 27.49/0.7754 \\ \cline{2-7} 
		& 2 & -/-        & 26.81/0.7512 & 27.59/0.7762 & \textcolor{red}{27.81}/\textcolor{red}{0.7851} & 27.43/0.7733 \\ \cline{2-7} 
		& 3 & -/-        & 26.70/0.7536 & 27.47/0.7720 & \textcolor{blue}{27.66}/\textcolor{blue}{0.7768} & 27.10/0.7577 \\ \hline
	\end{tabular}
	\label{tab:ablation}
\end{table}
\section{Ablation Study}

We evaluate different configurations of our network. For the results with limited training time and GPU memory, we only vary the number of hallucination maps in the SCH module between 0 to 3 and set the number of SCH modules to be 0, 1, 4, 8 and 12. As shown in Table~\ref{tab:ablation}, version 2-8 (2 maps and 8 modules) achieves the best performance and version 3-8 is the second best. Version 0-0 is the worst as it only contains one convolution layer and one resblock before the pixel-shuffle convolution module~\cite{shi2016real}. It shows that our SCH module plays a critical role in increasing the performance of SR and in reducing the parameters. Even the version 2-1 with only one SCH module can achieve almost the same performance as version 0-8 which contains over 10 convolution layers but without any hallucination maps. The performance becomes worse when we increase the number of hallucination maps beyond 2 and the number of SCH modules beyond 8. It may be due to overfitting with too many parameters. 

Meanwhile, in order to visualize the impact of the hallucination outputs, we also set 1 or 2 of the hallucination outputs (in all SCH modules) as 0. Note that we do this after the network is well-trained. As shown in Figure~\ref{fig:remove}, hallucination outputs are critical for the SOTA performance. In particular, the 1st hallucination output is more important for color tone restoration, and the 2nd is more important for deblurring and denoising. More visual comparison on the same datasets are shown in Section 4 and 6 of the supplementary materials.

\section{Conclusions}
In this paper, we propose a new spatial context hallucination network for blind SR tasks. To our best knowledge, we are the first one to propose such an idea. Our SCHN introduces an SCH module which simulates the procedure of multi-frame SR by generating pseudo frames, and achieves the SOTA performance comparing with existing SR methods. However, we have not combined internal (as in ZSSR) and external learning. Hence, our future work will focus on using internal information of the testing images. We also plan to train our network on a larger dataset to avoid overfitting. In this paper, we did not do this because we want to compare our model with others that are trained on the same datasets.

%References are listed in alphabetic order by the surname of the first author, or the identifying word (e.g., in case of a website). Have
%all anonymized references at the beginning of the list.

%here would be your acknowledgement (if any) in the final accepted paper

%===========================================================
\bibliographystyle{splncs}
\bibliography{Arxiv_submission}

%this would normally be the end of your paper, but you may also have an appendix
%within the given limit of number of pages
\end{document}